\documentstyle[12pt,english]{article}
\begin{document}

\begin{center}

{\bf Quantum spin liquid in  antiferromagnetic chain S=1/2 with
Acoustic Phonons }

\end{center}
\begin{center}
{\bf S.S. Aplesnin}
 \end{center}
\begin{center}
L.V. Kirenskii Institute of Physics, Siberian Branch of the
Russian Academy of Sciences, Krasnoyarsk, 660036, Russia , FAX 07
3912 438923, E-mail : apl@iph.krasn.ru
\end{center}

A spin and phonon excitations spectrum are studied using quantum
Monte Carlo method in antiferromagnetic chain with spins $S=1/2$
coupled nonadiabaticity with acoustic phonons . It is found  the
critical coupling exists to open gap in the triplet excitation
spectrum for any phonon velocity. The phase boundaries of
delocalized  phonons and propagated the bound states of magnon and
a phonon are calculated. It is shown that the spherical symmetry
of the spin-spin correlation functions is broken . The magnetic
and optical properties $CuGeO_3$ are explained without using
spin-Peierls transition.\\


The existence and stability of multiquanta bound states are now
established for a wide variety models with prescribed nonlinearity
and exhibited some distinctive observable signatures in terms of
spatiotemporal correlations. The typical origin of effective
nonlinearity in quantum system is the coupling of two or more
fields. Adiabatic slaving of fields usually results in nonlinear
Schrodinger models. Realistically nonadiabatic effects must be
considered and the influences of nonlinearity and nonadiabaticity
are inevitably interrelated. Here we consider spin-phonon coupled
model frequently used to describe the opening of a singlet-triplet
spin gap and an isotropic drop of the magnetic susceptibility
below critical temperature. As a rule a optical phonon interacts
with spin system and lead to lattice instability. The interaction
between acoustic phonons and spins has been studied in two limit
cases. A weak coupling lead to a renomalization of the phonon mode
and give rise to four spin interaction. As a result of a gap is
opened but the other well-defined magnetic and phonon bound states
has been loosed. A strong coupling dos't correspond to real
compounds and the results obtained in the intermediate coupling
are incorrect.

A magnetic, elastic and optical properties in the inorganic
quasi-one-dimensional  $CuGeO_3 $ compound fail to describe in
terms of one model. The thorough structure measurements
\cite{Loosd} and Cu-nuclear quadrupole resonance experiments
\cite{quadr} have not been found a lattice dimerization and a
softening of optical phonons. The static distortions below
critical temperature $T_c=14 K$ cannot be ascribed to any
pseudomode  in the high-symmetry phase \cite{Braden}. The magneto
induced structural effects depend on the  used frequency , so
strong lattice fluctuations have been observed by electron
diffractions up to $T \sim 100 K$ \cite{Chen}, by optical methods
up to $T \sim 200 K$ \cite{Braden2}. A frustration interaction due
to a asymmetric or a next nearest- neighbors antiferromagnetic
exchange can cause a gap in the triplet excitation spectrum though
no experimental proves have existence of these exchanges. The
physical nature of the critical temperature existence and opening
gap in triplet excitation spectrum, the  Raman spectrum, thermal
conductivity in magnetic
field remains unclear. \\
Numerical methods, such as exact diagonalization and
density-matrix renormalization group, face the potential
difficulty in dealing with a very large Hilbert space of the
phonons.  Here , we use Monte Carlo approaches restricted to
finite chains $L=100, 200$ but without any adiabatic approximation
and the truncation of the infinite phonon Hilbert space. The
method \cite{contin} is based on a path-integral representation
for discrete system in which we work directly in the Euclidean
time continuum. All the configuration update procedures contain no
small parameters . Being based on local updates only, it allows to
work with the grand canonical ensemble and to calculate any
dynamical correlation function, expectation values.\\
 We consider a model Hamiltonian of an  spin-phonon system:
\begin{equation}\label{1}
H=\sum_{i=1}^{L}[J+\alpha
(u_{i}-u_{i+1})][S_{i}^{z}S_{i+1}^{z}+(S_{i}^{+}S_{i+1}^{-}+S_{i}^{-}S_{i+1}^{+})/2]+M
\dot{u}_{i}^{2}/2+K(u_{i}-u_{i+1})^{2}/2.\;\;
\end{equation}
Here $S^{z,\pm} $ are a spin operator components associated with
the site i, $ J>0 $ is the usual antiferromagnetic exchange
integral, $\alpha$ is the spin-phonon coupling constant, $u_i$ is
the displacement in the z- direction, $M$ is the mass of the ion
and $K$ the spring constant. Using the quantum representation for
phonon operators $ b,b^{+} $, the Hamiltonian becomes
\begin{equation}\label{2}
H=\sum_{q}\sum_{i=1}^{L}[J+\alpha \sqrt{\frac{2\hbar }{M\omega _{0}}}\sqrt{%
\sin {\frac{q}{2}}}\cos {(q(i-0.5))}(b_{q}+b_{q}^{+})]
[S_{i}^{z}S_{i+1}^{z}+(S_{i}^{+}S_{i+1}^{-}+S_{i}^{-}S_{i+1}^{+})/2]+
\end{equation}
$$
\sum_{q}%
\hbar \Omega (q)b_{q}^{+}b_{q},\;\;\Omega (q)=2\sqrt{\frac{K}{M}}\sin {(%
\frac{q}{2})},\;\omega _{0}=2\sqrt{\frac{K}{M}},\;q=2\pi n/L,
n=1,2,...L
$$

The phonon frequency , spin-phonon coupling, energy, temperature
are normalized on the exchange $J$. The temperature used in
calculation is $\beta=J/T=25$. Our system  consists of the two
subsystem  interacted. The elastic subsystem is described by
phonons with the number of occupation $n_{ph}=0,1,2..$ and
magnetic subsystem is characterized by the number occupation
$n_{m}=0,1$ and Pauli operators $a,a^+$  which coincide with
$S^{\pm}$ spin operators.  We start with the standard Green
function of the phonon in the momentum $q$- imarginary-time $\tau$
representation:
\begin{equation}\label{3}
G(q,\tau
)=\sum_{\nu}|<\nu|b^{+}_q|vac>|^2\exp{[-(E_{\nu}(q)-E_{0})\tau]},
\end{equation}
where  ${|\nu>}$ is a complete set of the Hamiltonian $ H$ in the
sector of given $q$ , $ H|\nu(q)>=E_{\nu}(q)|\nu(q)>,\;
H|vac>=E_{0}|vac>, E_0=0$. Rewriting Eq.(3) as
\begin{equation}\label{4}
  G(q,\tau)=\int_0^{\infty} d\omega A(q,\omega) \exp{-(\omega
\tau)},\; \\
 A(q,\omega)=\sum_{\nu}
\delta(\omega-E_{\nu}(q))|<\nu|b^{+}_q|vac>|^2
\end{equation}
one defines the spectral function $A(q,\omega)$ . We calculate a
one-particle spin Green function
\begin{equation}\label{5}
  <a(\tau)a^{+}(0)>=\int_0^{\infty}d\omega \rho_t(\omega) \exp{-(\omega
\tau)},\\
\end{equation}
where  $\rho_t(\omega)$ -spectral density function , and
two-particle spin Green function associated with singlet
excitation $ \Delta S^z=0$
\begin{equation}\label{6}
  <a_0(\tau)a^{+}_1(\tau)a_1(0)a^{+}_0(0)>=\int_0^{\infty}d\omega
\rho_s(\omega) \exp{-(\omega \tau)}.
\end{equation}
Correlation functions are determined on the basis of a complete
set of eigenstates of the Hamiltonian:
\begin{equation}\label{7}
  <O>=\frac{\sum_{\nu} <\nu_i|O|\nu_j>}{\sum_{\nu} <\nu_i|\nu_j>},
\end{equation}
where $O=(a_0a^{+}_1+a_0^{+}a_1), (b_q+b^{+}_q)S_0^zS_1^z$,
$(a_0a^{+}_1+a_0^{+}a_1)(b_q+b^{+}_q) $. The longitudinal
spin-spin correlation function  $<S^z_iS^z_{i+h}>$, the phonon
density- density  $<n_{ph}(q)n_{ph}(q+p)>$, the distribution of
phonons number  $n_{ph}(q)$ and magnons number  $n_m(k) $ as a
function of momentum are simulated.\\
Energies excitation of the two-particles $S^+_kS^-_{k-p}, p=\pi$
and one- particle excitation differ by less than $ \sim 1 \% $ in
the isotropic antiferromagnetic chains. The interaction with the
elastic system lead to qualitatively various dispersion relation.
The two-particle excitation spectrum became gapped and
one-particle excitation gapless for $\alpha < \alpha_{c1} $. The
gap energy  determined from spectral density  well fit on the
linear dependence $\Delta_s(\alpha)\sim \frac{14\alpha}{\omega_0},
\; \Delta <J $ . For small  spin-phonon coupling $\alpha <
\alpha_{c1} $ the phonons are created near the upper bound of
triplet excitation band $W_t$ with momentum $q=2\arcsin{\frac
{W_t}{\omega_0}}$. It follow from the distribution of the phonon
number versus momentum and the spectral density. The spins number
in the structure of the spin-phonon quasiparticle
$b^+_{q-k_1-k_2-...-k_n}S^{\gamma}_{k_1}S^{\gamma}_{k_2}...S^{\gamma}_{k_n}
, \gamma=z, \pm$ increase and the correlation function between
spin and phonon
\begin{equation}\label{8}
  R_{bs}=\sqrt{\frac{2\hbar}{M}}<\sqrt{\sin{\frac{q}{2}}}\cos{(q(i-0.5))}(b^+_q+b_q)S^z_iS^z_{i+1}>,
\end{equation}
qualitatively confirms  it as illustrated in Fig.1a . For
$\alpha<\alpha_{c2}$ the correlation function is parameterized as
 $R_{bs}\simeq
\frac{\alpha}{2\omega_0}\; ,\omega_0>3$ è $R_{bs}\simeq
\frac{\alpha}{2W_t}\;, \omega_0\leq3 $. Correlation function
between the longitudinal acoustic phonon and the transverse spin
components $<(b^++b) a^+_0 a_1> $  is less  by a factor 3-5 than
$R_{bs}$. Number of two-particle singlet excitation falls as to an
exponent $N_s \simeq
\frac{1}{4}\exp{-[\frac{3\alpha}{2\alpha_{c2}}]}$. The formation
of the bound states and nonlinear excitations causes to break the
spherical symmetry of the spin-spin correlation function. The
ratio of the correlation functions of transverse spin components
to longitudinal is parameterized as the linear dependence on the
spin-phonon coupling
\begin{equation}\label{9}
  [\frac{<S^+_0S^-_1>}{<S^z_0S^z_1>}-2]\simeq
  \frac{5,2\alpha}{\omega_0} ,\; \alpha<\alpha_{c2} ,
\end{equation}\\
as presented in Fig.1b.
 The magnetic susceptibility is represented by spin-spin correlation
function when the long-range order is absent $
\chi_{x(y)}/\chi_{zz}=\sum_r{S^+_0S^-_r}/\sum_r{S^z_0S^z_r}.$
 Conventionally the  physical nature of anisotropy result from spin-orbital
 interaction for spin $S=1/2$. We suggest a new mechanism
the dynamical interaction between the longitudinal acoustic phonon
and spin. It is important for S- ions.

For   $\alpha > \alpha_{c1}$ the gap in the triplet excitation
spectrum simulated from the spectral density function is open .
The functions $\rho $ are  plotted in Fig.2. The gaps energies
 fit satisfactory on the straight line $\Delta
\simeq 0.8(9) (\alpha-\alpha_{c1}) $ in the range of
$\alpha_{c1}<\alpha <\alpha_{c2}$. Here exists the bound states of
spin-phonon quasiparticls similar to bipolaron with the symbolic
kind $<b^+_{\textbf q-\sum_i^{n_1} k_i}b^+_{\textbf
q+p-\sum_i^{n_2} k_i} \prod_i^{n_1+n_2} S^z_{\textbf k_i }> $. The
correlation function of the phonon number
  $<N_{ph}(q)N_{ph}(q+p)>$, presented in Fig.3b is nonzero at the certain
  wave number  $p$
  for   $\alpha \simeq \alpha_{c1}$ and for all $p$ at the condition  $\alpha > \alpha_{c2}$.
The magnon distribution function  demonstrates a small
oscillations approximately within a five percents as compared to
$n_{m}(q=0,\pi)\simeq 0.22 $ for $\alpha
> \alpha_{c1}$ and the number phonon distribution function  reveals four local maxima for $\alpha
> \alpha_{c2} $ and one sharp maximum for $\alpha
> \alpha_{c3}$.  Analysis of the spin- spin correlation function
plotted in Fig.3 lead to two important length scales:
\begin{equation}\label{10}
  |<S^z_0 S^z_r>| \simeq \frac{A}{r} \exp{(-r/\xi)},\; r< r_c, \\
 <S^z_0 S^z_r> \simeq (-1)^r  \frac{\cos{(Qr)}\cos{(qr)}}{r^{\theta}}, \; r > r_c.
\end{equation}
 where  $\xi$ correlation radius,  $Q, q, \theta $ fitted parameters.
$R \sim \pi/Q$ may be interpreted as the typical size of a cloud
of spins and phonons ; $\xi$ may be considered as the width of the
transition region separating the nearest spin-phonon
quasiparticles. These coherent nonlinear excitations cause the
local extremums of the spectral density phonons and spin
excitations at the same energies  including the spin gap range as
shown in Fig.2. It is possibly a standing vibration with the spin
kink in sites exists here. The  disjoint  vibrations are performed
at the condition that the wave length is changed in two times
$l/a=2; 4; 8; 16; ...$ c $k=\pi/l, E=w_0 sin(k/2)$. These
estimates of energy are in good agreement with Monte Carlo results
up to $n=4$ because the temperature  fluctuations  $(T=0.04), \xi<
l/a=64$ cut the kinks interaction radius. The average number of
phonons increase vs. $\alpha$ according to the power law
$$N_{av}=0,0012
(\frac{\alpha}{\alpha_{c2}})^{1.75(6)}, v_{ph} \le v_m; \;
N_{av}=0,0006 (\frac{\alpha}{\alpha_{c2}})^{2.4(6)}, v_{ph} >
v_m,\; (11)
$$
here $v_{ph}$ and $v_m$ is the velocity of phonon and magnon. The
spectral density of phonons and the distribution of phonon number
became continuum at $\alpha=\alpha_{c2}$. The value $\alpha_{c2}$
is similar to critical concentration when dressed phonons
effectively "percolate " along the three low-lying bands. The
correlation function $<N_{ph}(q)N_{ph}(q+p)>$ is not equal to zero
for all momentums as shown in Fig.3. The transition from one state
with localized phonons to delocalized is like to Anderson's
transition in disordered systems. The width of band increases
versus $\alpha$ as
 $W_{ph} (\alpha_{c2})/ W_{ph}(\alpha_{c1}) \sim 1.5 $.
  When the spin-phonon coupling is compared to the
energy of the  boundary of top band of phonon and triplet spin
excitation $\alpha
> \alpha_{c3} \sim 0.8 w_0 \;$ the soliton lattice is formed with
the wave vector of structure being in the range of $\pi < Q < 2
\pi$ . A nonlianear excitation ,for example, the spin breather is
gapless and the  spectral density of phonons and   spin
excitations  are comparable values   at the low energies . The gap
in one-particle spectrum excitation pass  into quasigap. The area
of parameters in the plane the energy of top band of acoustic
phonon  $\omega_0$ - the value of the spin-phonon coupling $
\alpha $
  , where a gap in the one-particle excitation
 spectrum is arisen, is limited by two fitting lines $\alpha_{c1}
\simeq \frac{\omega_0}{4 W_J}$ and $\alpha_{c3} \sim 0.8 w_0 $ .
The complete filling of the three acoustic bands occurs at the
$\alpha_{c2} \simeq \frac{\omega_0}{2 W_J}$.

Now we will apply our results for interpretation and prediction of
new effects for the one-dimensional spin system $ CuGeO_3 $ with
intrachain  exchange  $J_{Ge}=120 K,  J_1/J_{Ge}=0.1 $
\cite{Nishi}. The crystal structure $ CuSiO_3 $ is isostructural
to $ CuGeO_3 $. However $ CuSiO_3 $ is antiferromagnet with $
J_{Si}=21 K, J_1/J_{Si}=0.14 $ ($J_1 -$ interchain  exchange
)\cite{Baenitz}. The interaction between  spins and  acoustic
phonon mode in chain leads to gap in the triplet excitation
spectrum $ CuGeO_3 $. For the same value of the spin - phonon
coupling is $[\omega_0/J]_{Si}>> [\omega_0/J]_{Ge} $ and
$\alpha_{c1,Si}
> \alpha_{c1,Ge} $. The magnon velocity  $ CuGeO_3$ is $ v_m\sim
1300 m/s $, the phonon velocity depends on the ultrasound
frequency \cite{Panl} and we take the average value
 $v_{ph} \sim 5000 m/s $  è $ \frac{v_{ph}}{v_m}=\frac{\omega_0}{\pi
 J}=3,84 $. The anisotropic elastic constants are $ c_{11}=64 GPa, c_{22}=37.6 GPa, c_{33}=317.3 GPa
   $ \cite{Ecol} and confirm validity using one dimensional model
  to study the dynamic of the interacting magnetic and elastic
  systems. According to our results the spin-phonon coupling in $ CuGeO_3$
  has been established   $\alpha \simeq 2,2$ for $\Delta/J=0.2 ,\omega_0=12 $.
  The estimated energies of the bound spin-phonon excitation are $E_{sph}=95 cm^{-1} , 48 cm^{-1}, 24.4 cm^{-1}, 12.5 cm^{-1}, 6.2 cm^{-1},
3 cm^{-1}$ and are in general agreement with Raman spectrum $
E_{sph}^{ex}=98 cm^{-1}, 48 cm^{-1}, 30 cm^{-1}, 13.2 cm^{-1} $
\cite{Raman}. The localized phonon excitation in the second
$E^{MC}_2=0.22 eV $ and in the third $ E^{MC}_3=0.34 eV $ bands
agree satisfactory with data of the optical absorbtion $E^{ex}_2
=0.21 eV $ \cite{Devic},
 $ E^{ex}_3=0,36 eV $ \cite{Devic1} and make clear the physical origin
 of these peculiarities . Up to present time the existence of isolated energy
 level $E^{ex} \sim 6 cm^{-1} $ below the triplet gap has been remained unclear. This gap is found by
  the inelastic neutron  scattering \cite{Ain} and the submillimiter
  resonance \cite{Smirnov}. We predict a new resonances at the frequencies $\omega \sim 130 GHz; \sim 65 GHz $
  at  low  temperatures  $T< 1 K$.

  The bound spin-phonon   excitations will result in additional phonon scattering which
  dependents on magnetic field.  The energy  of spin-phonon
  quasiparticle normalized on gap value  $E^{MC}_i/\Delta=0.37;
  1.5$ is well agreement with temperatures normalized on critical
  temperature $T_c=14 K$ ,$T^{ex}/T_c \simeq
0.39; 1.57 $ related to local maxima of thermal conductivity
\cite{Ando}. This effect result from interaction between elastic
and magnetic system and  magnetostriction parameter exhibits also
two maximum at the same temperatures \cite{magnit}. The
relationship of transverse spin-spin correlation function to
longitudinal  $[\frac{<S^+_0S^-_1>}{<S^z_0S^z_1>}-2]\simeq 0,9$ is
associated with the anisotropy exchange $(J_b/J_c)^{MC}=1,45$
which agrees  with experimental data
$J_b/J_c=\Theta_b/\Theta_c=1,38$
  \cite{Petrak}. The sharp rise of $g-$ factor near the transition
  $T>  T_s$  may be explained in term of formation
  of the coherent state of spin-phonon quasiparticles which may be
  broken by thermal phonons. The average number of phonons
  resulted in spin-phonon interaction   $<N_{0}>=1,7\cdot 10^{-4}$ in $CuGeO_3$
  is  equal to number of thermal phonons define in terms of Debye
  approximation with Debye temperature $\Theta=330 K$ \cite{Debye}
  at the  temperature $T=18 K$.

   In conclusion, we summarize  the main results.  The interaction
   between magnetic and elastic system lead to gap in one-particle
   and two-particle excitation spectrum. The gapped triplet
   spectrum arises at the certain critical value of the
   spin-phonon coupling. Bound spin-phonon excitation localize for
   $\alpha< \alpha_{c3} $ and causes a  peculiarities set of
   spectral density of phonon and spin excitations. Spherical
   symmetry between transverse and longitudinal  spin-spin
   correlation function is broken. The triplet gap in $CuGeO_3$
   arises from the interaction between spin and acoustic phonon and
   this compound has not possessed spin-Peierls  ordering. From
   our results the magnetic , elastic and optic properties $CuGeO_3$
  is explained.

   The author are grateful to V.A. Kashurnikov, A.S. Mishchenko,
   N.B. Prokof'ev, B.S. Svistunov for useful assistance and
   discussions of Monte Carlo method in the continuous time.

\newpage
Captions to Aplesnin paper ñ S=1/2 " Quantum spin liquid in
antiferromagnetic chain S=1/2 with Acoustic Phonons"

Fig.1 The correlation function of the nearest-neighbor spin-spin
and displacement $<A_{qi}(b^+_q+b_q)S^z_iS^z_{i+1}>$, where
$A_{qi}=\sqrt{\frac{2\hbar}{M}}\sqrt{\sin{\frac{q}{2}}}\cos{(q(i-0.5))}$
 $(a)$ and ratio transverse spin-spin correlation
function to longitudinal  $(b)$ with  $\omega_0=10 (1), 8(2),
6(3),  1(4)  $
 versus  spin-phonon coupling . \\
 Fig.2  The spectral density one-particle excitations with $\omega_0=8,
 \alpha=0.4 (1), 2.6(2), 3.2(3) (a)$, $ \alpha=1.6 (1), 2 (2) (b);
 $ and two-particle singlet excitations with
  $\omega_0=8, \alpha=2.6(2), 3.2(1) (c) $ and phonon excitations
  with $\omega_0=8,  \alpha=2(1), 3.2(2) (d)$ .\\
Fig.3 Longitudinal spin-spin correlation function vr. $r$ with
  $\omega_0=8, \alpha=0.8 (1), 2 (2), 3.2(3)(a) $.
  Correlation function of the phonon number normalized on its maximum
  value versus momentum with  $\omega_0=8, \alpha=1.4
(1), 2 (2), 2.6(3)   (b). $ \\

$\vert$%

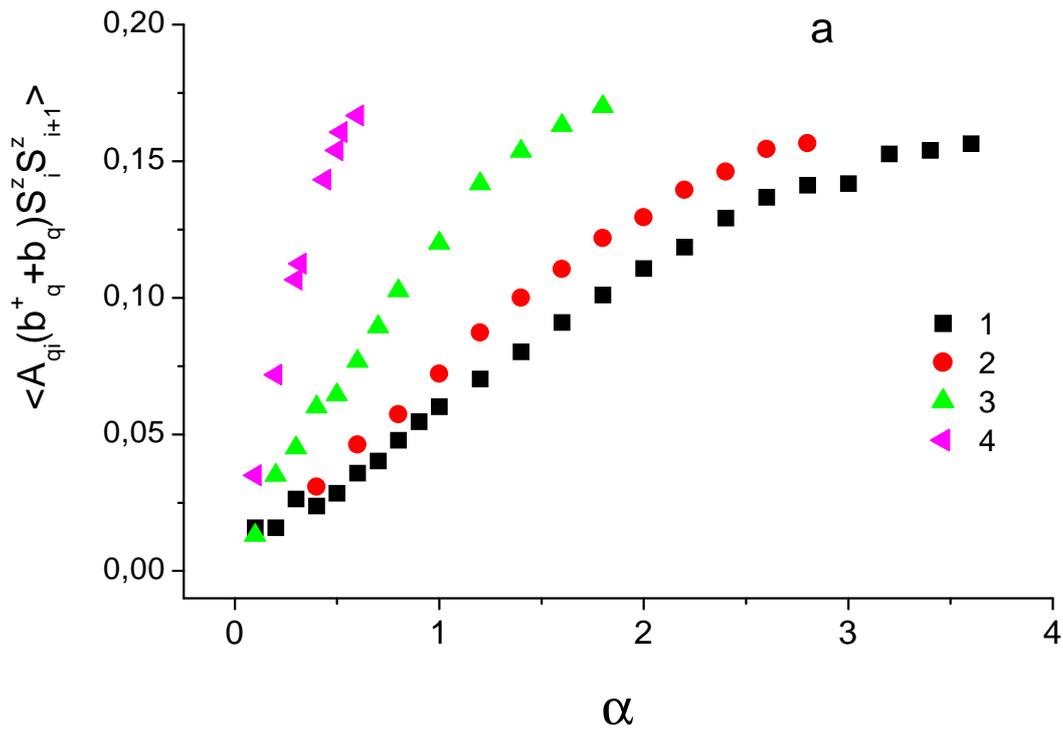
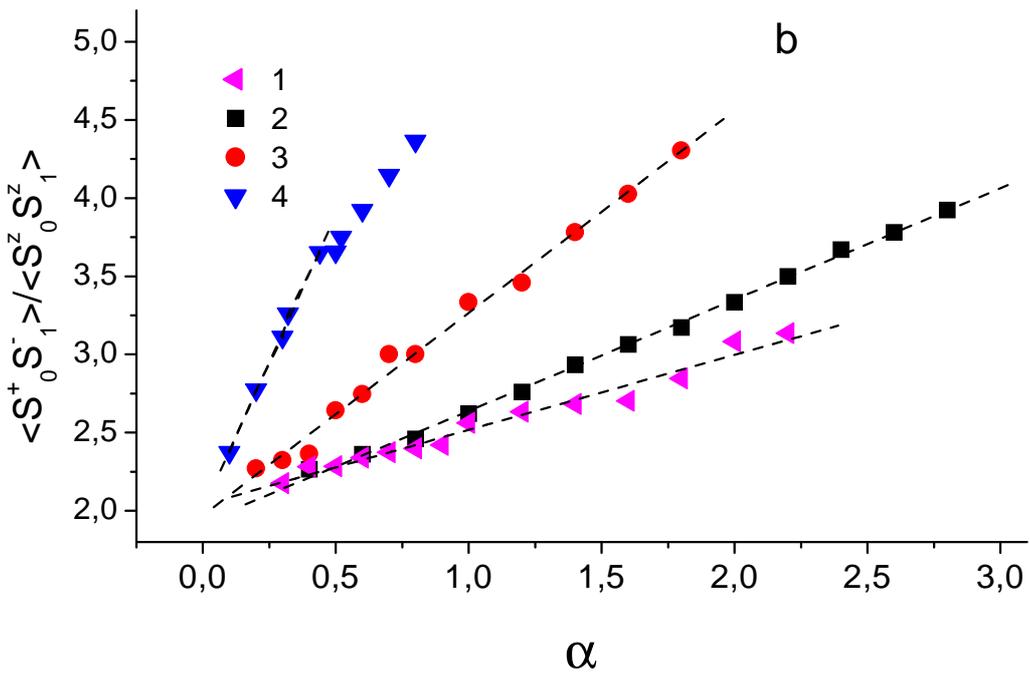
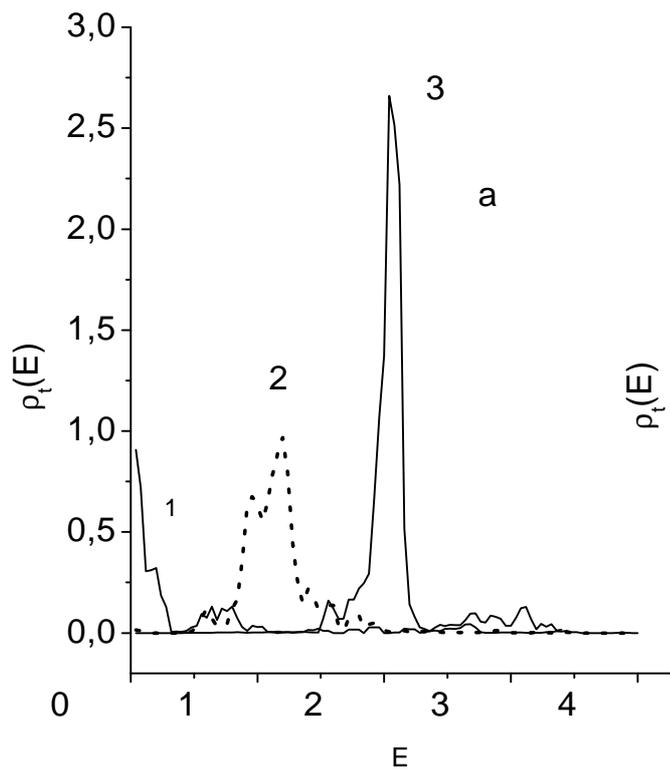
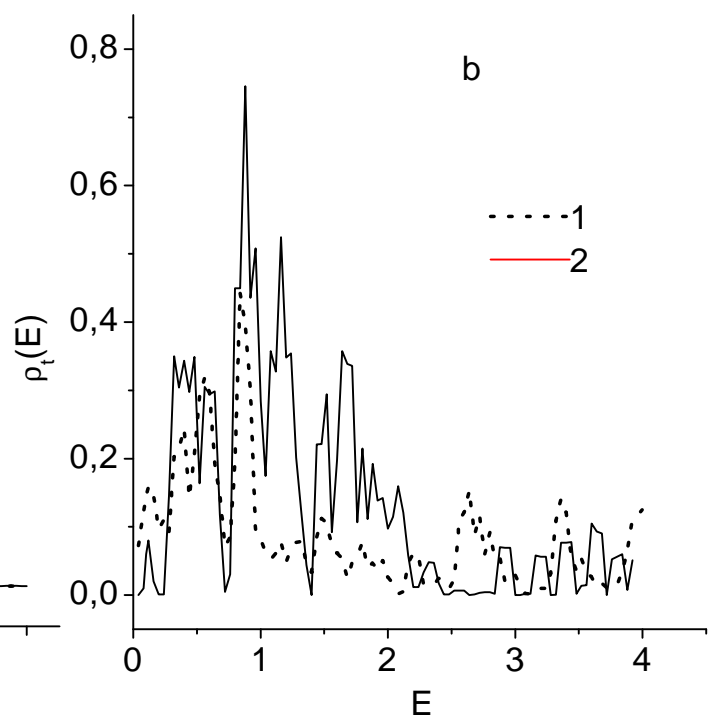
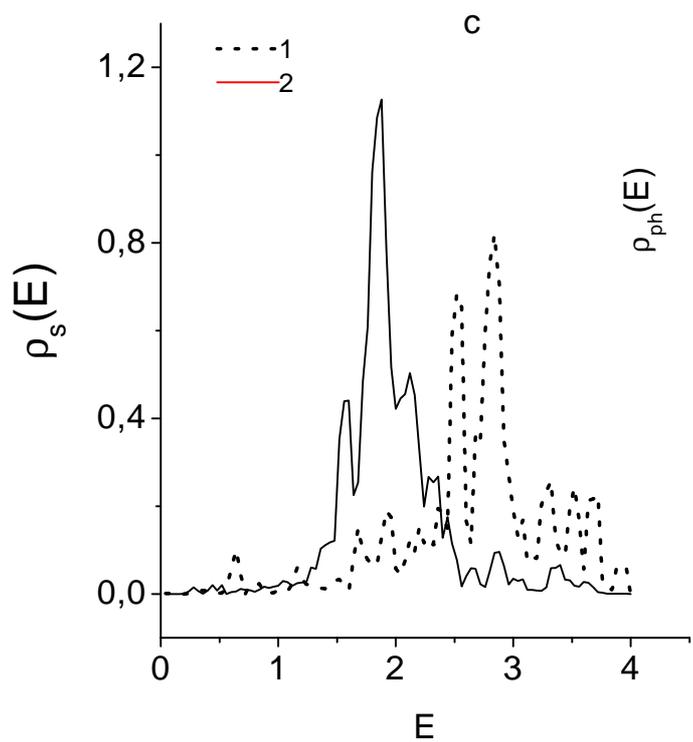
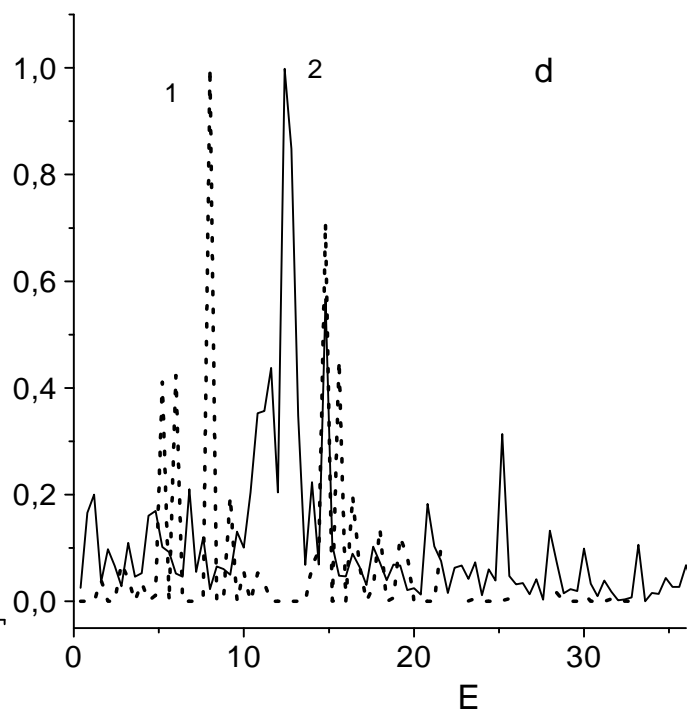
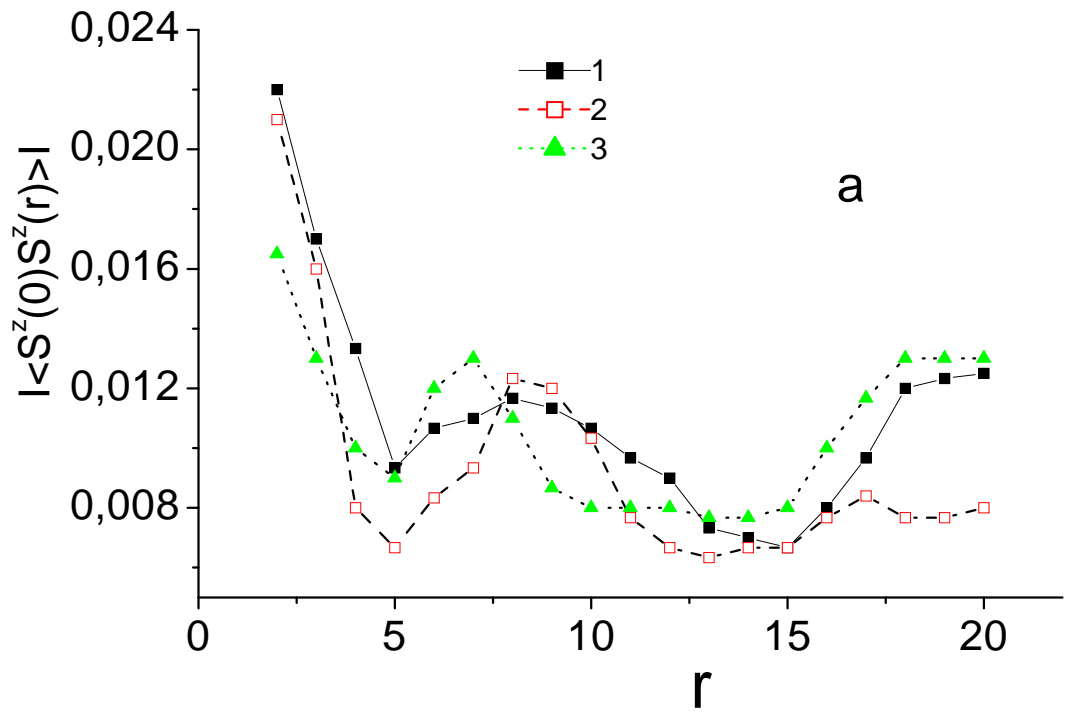
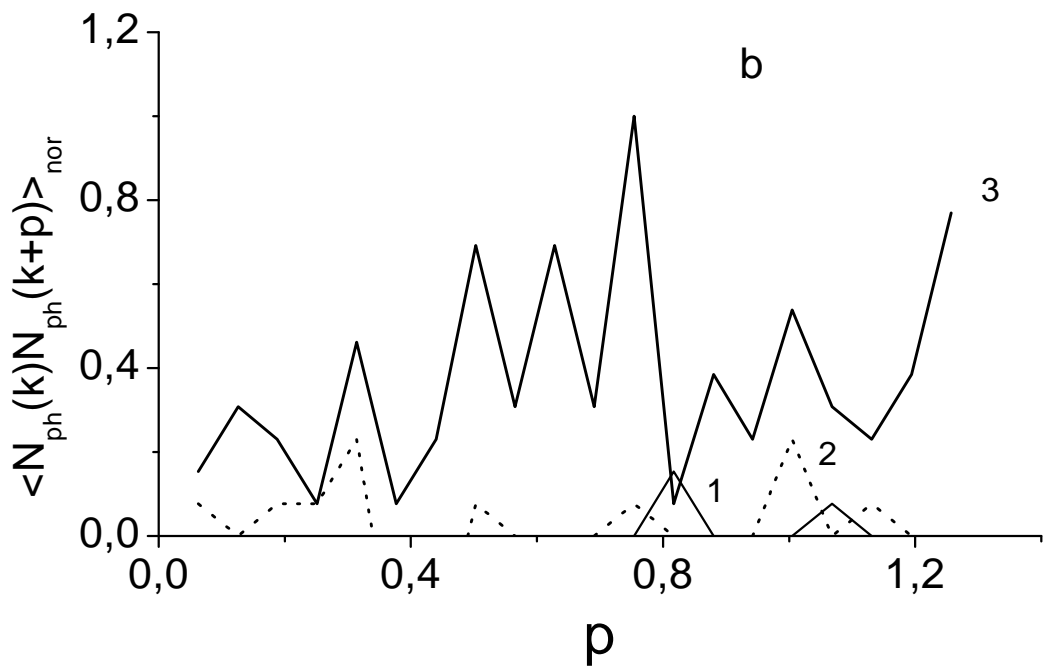
\end{document}